\documentclass[a4paper,fleqn,usenatbib]{mnras}
\usepackage[T1]{fontenc}
\usepackage{ae,aecompl}

\usepackage{color}
\usepackage[dvipsnames]{xcolor}

\newcommand{\nucl}[2]{\mbox{$^{#1}${#2}}}

\newcommand{\vunit}{\mbox{\,km\,s$^{-1}$}}

\newcommand{\Msun}{\mbox{\,$M_\odot$}}
\newcommand{\Lsun}{\mbox{\,$L_\odot$}}
\newcommand{\Rsun}{\mbox{\,$R_\odot$}}

\newcommand{\mic}{\mbox{$\,\mu$m}} 
\newcommand{\us}{\mbox{U~Sco}}

\newcommand{\fion}[2]{[\ion{#1}{#2}]}

\newcommand{\ltsimeq}{\raisebox{-0.6ex}{$\,\stackrel 
	{\raisebox{-.2ex}{$\textstyle <$}}{\sim}\,$}} 
\newcommand{\gtsimeq}{\raisebox{-0.6ex}{$\,\stackrel
	{\raisebox{-.2ex}{$\textstyle >$}}{\sim}\,$}}

\usepackage{graphicx}	
\usepackage{amsmath}	
\usepackage{amssymb}	

\title[NIR spectroscopy of LMCN 1968-12a]{Near-infrared spectroscopy of the LMC recurrent nova LMCN 1968-12a}

\author[A. Evans et al.]{A. Evans,$^{1}$\thanks{E-mail: a.evans@keele.ac.uk},
D. P. K. Banerjee$^{2}$,
T. R. Geballe$^{3}$,  
A. Polin$^{4}$, 
E. Y. Hsiao$^{5}$,  \newauthor
K. L. Page$^{6}$, 
C. E. Woodward$^{7}$,
S. Starrfield$^{8}$  
\mbox{ } \\ \\
$^{1}$Astrophysics Group, Lennard Jones Laboratory, Keele University, Keele,
Staffordshire,  ST5 5BG, UK\\ 
$^{2}$Physical Research Laboratory, Navrangpura,  Ahmedabad, Gujarat 
380009, India\\ 
${^3}$Gemini Observatory/NSF's NOIRLab, 670 N. Aohoku Place, Hilo, HI, 96720,
USA\\  
$^{4}$Department of Physics and Astronomy, Purdue University, 525 Northwestern Avenue, West Lafayette, IN 47907, USA \\ 
$^{5}$Department of Physics, Florida State University, 77 Chieftan Way, Tallahassee, FL 32306, USA\\
$^{6}$School of Physics and Astronomy, University of Leicester, University Road, Leicester, LE1 7RH, UK \\ 
$^{7}$Minnesota Institute for Astrophysics, School of Physics \& Astronomy,
116 Church Street SE, University of Minnesota, \\
Minneapolis, MN 55455, USA\\ 
$^{8}$School of Earth and Space Exploration, Arizona State University, 
Box 876004, Tempe, AZ 85287-6004, USA\\ 
}

\date{Accepted XXX. Received YYY; in original form ZZZ}

\pubyear{2024}

\begin{document}
\label{firstpage}
\pagerange{\pageref{firstpage}--\pageref{lastpage}}
\maketitle

\begin{abstract}
We have obtained near-infrared  (0.80--2.45\mic)
spectra of the recurrent nova
LMCN 1968-12a on two occasions during its 2024
August eruption. This is the first near-infrared spectroscopy
of an extragalactic nova. The initial spectrum, on day 8.48, 
caught the nova in the coronal phase, with the
[\ion{Si}{X}] 1.43\mic\ line being extremely strong.
This line had a luminosity of $\sim95$\Lsun, and is clearly a
very powerful coolant. Its presence, together with the absence of
[\ion{Si}{ix}] 1.56\mic, implies a coronal temperature
$\gtsimeq3\times10^6$~K, possibly amongst the highest recorded
coronal temperature in a nova eruption.
 With the exception of the [\ion{Si}{X}] line,
the near-infrared spectra are remarkable for being devoid of metal lines.
We suggest that this is due, in part, to the exceptionally
high temperature of the coronal gas, causing ions, whose emission
lines would normally appear in the near-infrared spectrum, to be
collisionally ionised to higher stages.
\end{abstract}




\section{Introduction}
Nova explosions occur in semi-detached binary systems containing
a late-type star (the secondary) and a white dwarf (WD). 
Material from the secondary spills onto the WD via an accretion
disc. In time, conditions at the base of the layer accreted on
the WD become degenerate, and hot enough to trigger a thermonuclear
runaway (TNR). This produces a nova eruption, leading
to the violent ejection of material,
at several hundred to several thousand \vunit. Once the eruption has subsided, 
acc\-retion resumes and in time, another eruption occurs: 
all novae are recurrent. However the eruptions of {\em recurrent}
novae (RNe) repeat on time-scales $\ltsimeq100$~yr 
\citep[see][for reviews]{anupama08,darnley20,darnley21}.

RNe are defined by the selection effect that the system
has been observed to have undergone more than one TNR. 
Fewer than a dozen Galactic RNe are known. 
There are rather more extragalactic RNe;
the majority of these are in M31. There are also four
in the Large Magellanic Cloud (LMC) \citep{healy24}.
The second eruption in the LMC of ``Nova Mensae 1968'' in 1990 
\citep{liller90} meant that this object was the first 
extragalactic RN to be observed; its
most recent eruption occurred in 2024 August.

We present here near infrared (NIR) 
0.8--2.5\mic\ spectroscopy of the 2024 eruption,
the first NIR spectroscopic study of any extragalactic RN.
A preliminary account was given by \cite{banerjee24}.

\section{The LMC system}
\label{system}
Previous eruptions of Nova LMC 1968 were observed in 1968, 
1990 \citep{shore91}, 2002, 2010 \citep{mroz14}, 2016
\citep{mroz16,kuin20}, and 2020. A comprehensive discussion
of these eruptions is given by \cite{kuin20}, who 
concluded that it is unlikely that other eruptions
had occurred since 1999. The nova 
is known in the literature as LMCN 1968-12a,
LMCN 1990-02a, LMCN 2010-11a, LMCN 2016-01, Nova LMC 1990b, 
OGLE 2016-NOVA-1, and LMC V1341. We refer to it here as LMC68.

LMC68 has an orbital period of 1.26432~d \citep{mroz14},
suggesting that it is a RN of the ``\us'' type 
\citep[e.g.][]{anupama08}. The similarities between
LMC68 and \us\ have been discussed by \cite{kuin20}.
Unlike RNe with red giant (RG) secondaries, there is
no RG wind in \us-type RNe with which the ejected material
interacts. The comparable orbital periods
suggests that the secondary in LMC68
is similar to that in the \us\
system, in which the secondary is a sub-giant \citep{anupama08}. 

\cite{kuin20} give the $VI$ magnitudes in quiescence from the 
OGLE survey \citep{mroz14} as $\langle{V}\rangle=19.70$,
$\langle{I}\rangle=19.29$. Assuming a range
of $\pm0.5$ in these values \citep[see Figure~5 of][]{mroz14},
these data are 
consistent with a black body with temperature 
$\sim6700$~K and luminosity $\sim30$\Lsun,
although in view of the paucity of, and nature of, the data,
these values are subject to significant uncertainties
(see Fig.~\ref{cont23} below).

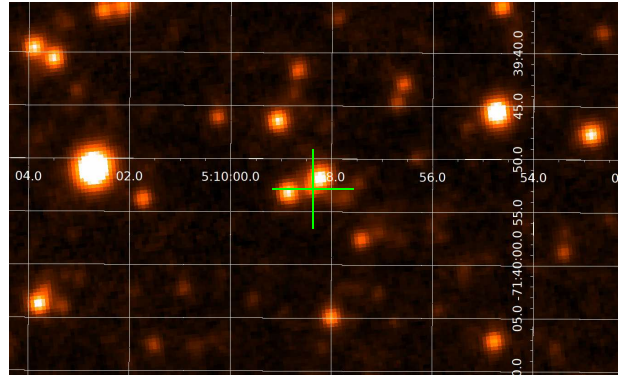
\begin{figure}
\begin{picture}(100,100)
 \includegraphics[width=8cm]{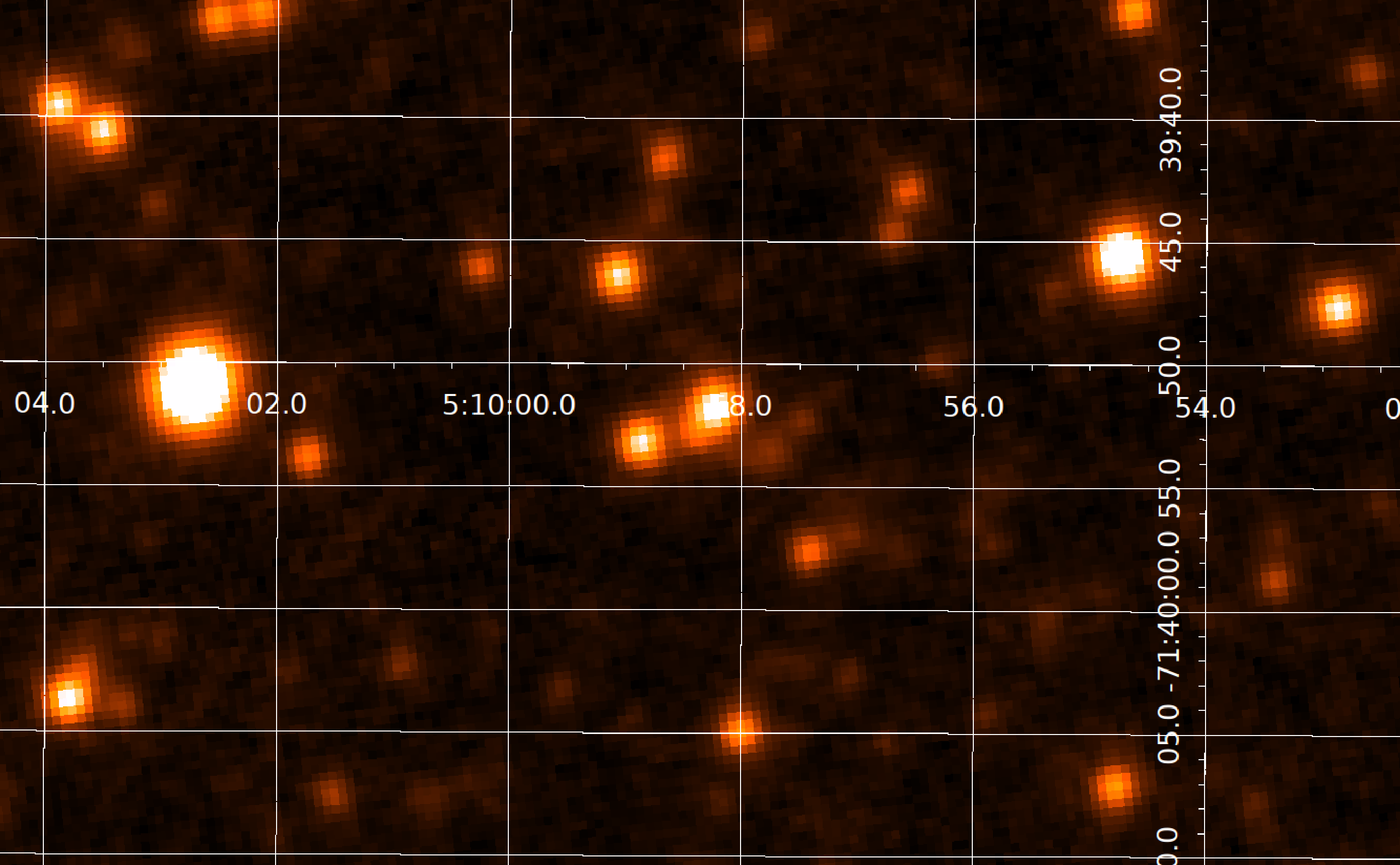} 
 \put(-129,70){\color{green}\line(1,0){30}}
 \put(-114,55){\color{green}\line(0,1){30}}
\end{picture}
 \caption{NIR field of LMC68. $Y$ image from the VISTA survey.
 LMC68 is located at the centre of the green cross;
 it is $\sim1\farcs5$ SE of a brighter star. 
 \label{vista}}
\end{figure}

There appear to be no deep NIR images of the nova field,
and it is not resolved in the Two Micron All Sky Survey
\citep[2MASS;][]{skrutskie06}. However, it was observed 
in the $Y,J$, and $K_s$ bands by the Visible and Infrared Survey
Telescope for Astronomy \citep[VISTA;][]{cioni11}
in 2014 November, when the system was in
quiescence. We reproduce the $Y$ band image in 
Fig.~\ref{vista} as it provides a useful finder
for NIR observations.

We assume the distance and reddening to LMC68 to be 50~kpc
\citep{pietrzyski19}, and $E(B-V)=0.07$ \citep{kuin20} 
respectively.

\section{The 2024 eruption}

Following the 2020 eruption, a recurrence period of four years
was proposed \citep*{page20}. Right on cue, LMC68 erupted
in 2024 August \citep*{darnley24,stubbings24}. 
The Neil Gehrels Swift Observatory \citep{gehrels04}
had been monitoring LMC68 with monthly cadence since the end
of the 2020 eruption. A Swift observation comprised of two
``snapshots'' taken some
16~hrs apart, was obtained on 2024 August~1. 
LMC68 was still in quiescence during the first
snapshot (obtained on August 1.19 UT), but clearly in eruption
during the second (August 1.83~UT). The 2024 eruption therefore
occurred between (UT) August 1.19 (MJD 60523.19) and August 1.83
(MJD 60523.83). We take zero of time from the latter,
$t_0=\mbox{MJD~}60523.83$.

Optical spectroscopy covering the wavelength range 3600--7800\AA\
was obtained on day~4.7 \citep{shore24}, when
\ion{He}{II} 4686\AA\ dominated the spectrum.
H$\alpha$ had broad wings extending to $\pm5000$\vunit.
Most interestingly, \citeauthor{shore24} note that there
were no metal lines present. They also found that the emission
line profiles differed from those presented by \cite{kuin20},
suggesting that the ejecta morphology may differ from eruption
to eruption, or that there may be orbital phase
effects \citep{kuin20}.

Swift observations obtained on day~7.8 revealed a soft X-ray
source \citep*{page24}, which was approximated 
by black body emission with temperature $T_{\rm Swift}$ given by 
$kT_{\rm Swift}=71^{+28}_{-50}$~eV; a deeper observation 
on the following day gave $kT_{\rm Swift} = 66\pm12$~eV.

\section{Observations}

\subsection{Magellan}
A low resolution spectrum covering the wavelength range 
0.8--2.5\mic\ was obtained on 2024 August 10.41 UTC
(MJD 60532.41 = day~8.58) at a mean airmass of 1.502, with the 
Folded Port Infrared Echellette (FIRE) spectrograph
on the 6.5~m Magellan Baade Telescope \citep{simcoe08}.
The observation consisted of standard
ABBA loops on LMC68, followed immediately by an
observation of the telluric standard HD\,223296.

The resolving powers, $R$, were 500 in the $J$ band,
450 ($H$), and 300 ($K$). The spectrum was reduced using a
custom IDL package \citep{simcoe08}
and additional procedures described by \cite{hsiao19}.

\subsection{Gemini-S}

NIR spectra covering 0.90--2.45\mic\ were obtained at Gemini 
South on 2024 August 24.32 (MJD 60546.318 = day~22.49). 
The facility instrument Flamingos-2 
\citep{eikenberry04} was used and was configured with its
0\farcs72 wide slit and its $J\!H$ (covering 0.9--1.75\mic)
and $H\!K$ (1.54--2.45\mic) grisms. With this slit
width the resolving power of the $J\!H$ grism varies
from $\sim200$ to
$\sim~700$ across its wavelength range, and that of the $H\!K$
grism varies from $\sim300$ to $\sim700$ across its wavelength
range\footnote{See https://www.gemini.edu/instrumentation/flamingos-2/ components\#Grisms.}. 
Spectra were obtained in the standard ABBA mode with a nod angle
of $6''$. Total exposure times were 1,440 seconds and 2,400
seconds for the short and long wavelength segments, respectively. 
The time intervals for the observations were UT 
06.99--07.47~hr for $J\!H$ and UT 07.74--08.55~hr for $H\!K$. 
The mean airmass was 
1.70 for the $J\!H$ segment and 1.55 for the $H\!K$ segment. 
The A0V stars HIP\,18842 and HIP\,30360 were employed as 
telluric standards for $J\!H$ and $H\!K$, respectively. 
The mean times of the (short) telluric standard
observations were UT 06.43 ($J\!H$) and UT 08.97 ($H\!K$).
Data reduction
incorporated both IRAF and Figaro routines as described in 
\cite{evans22}. Because of large difference in the OH sky 
line emission line brightnesses between the A and B spectra 
during the $H\!K$ measurements, it was necessary
to scale the $H\!K$ spectra in the A position by a factor of 0.9
before combining them with the spectra in the B position. 
This procedure removed the residual sky lines. 
The $H\!K$ spectrum was then scaled to match the $J\!H$ 
spectrum in their overlapping
wavelength interval (1.50--1.75\mic) and
both spectra were resampled 
in bins of 0.002\mic, prior to adjoining them. 

\begin{figure*}
 \includegraphics[width=12cm]{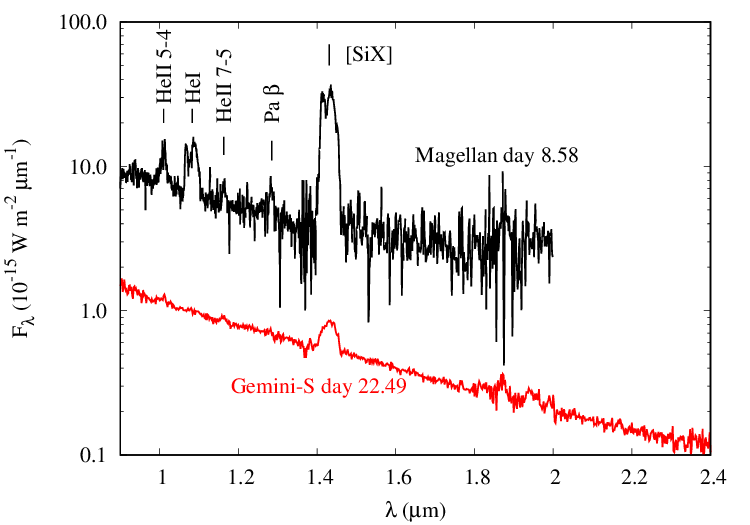}
 \caption{NIR spectra of LMC68, obtained on days 8.58 (black)
 and 22.49 (red). The apparent emission features
 around 1.8--2.0\mic\ are telluric in origin.\label{data}}
\end{figure*}

The Magellan and Gemini spectra are shown in Fig.~\ref{data}. 
All wavelengths are vacuum values.
The Magellan (day~8.58) data are shown up to 2.0\mic\ only,
as the data longward of this have low signal-to-noise ratio
($\ltsimeq3$).

\subsection{Swift}

LMC68 was observed with the Swift X-ray telescope (XRT) at
times close to  those of the NIR observations:
\begin{enumerate}
 \item on MJD 60533.692 (day~9.86), the data are fitted by a 
super soft X-ray source with temperature 
$kT_{\rm Swift}=49^{+11}_{-8}$~eV,
$T_{\rm Swift}=5.7^{+1.3}_{-0.9}\times10^5$~K, and
\item on MJD 60546.420 (day~22.59), with 
$kT_{\rm Swift}=122\pm10$~eV,
$T_{\rm Swift}=1.41\pm0.12\times10^6$~K.
\end{enumerate}
Following the discovery of the 2024 eruption by
Swift, a daily monitoring campaign was initiated. The data were
processed and analysed using {\sc HEASOFT} 
v6.34\footnote{https://heasarc.gsfc.nasa.gov/docs/software/heasoft/}, together with the most recent calibration files available at the time.

LMC68 was also observed with the UV/Optical telescope UVOT. 
The UVOT magnitudes are listed in Table~\ref{uvot},
together with the corresponding fluxes based on the
calibration given in \cite{breeveld11}.
The entire Swift dataset will be published elsewhere.

\begin{table*}
\caption{Observed UVOT fluxes$^*$. \label{uvot}}
 \begin{tabular}{ccccc} \hline
Filter & $\lambda$ & MJD & Mag & Flux \\
    & (\AA)     &   &  & ($10^{-14}$ W~m$^{-2}$\mic$^{-1}$) \\ \hline
$v$    & 5410 & 60533.0979 & $16.694\pm0.093$  &$0.78\pm0.07$ \\
$b$    & 4321 & 60533.0925 & $16.752\pm0.054$  & $1.29\pm0.06$ \\
$u$    & 3442 & 60533.0914 & $15.569\pm0.044$  & $2.09\pm0.08$ \\
$w1$   & 2486 & 60533.0897 & $15.295\pm0.041$  & $3.03\pm0.11$ \\
$m2$   & 2221 & 60533.0998 & $15.067\pm0.045$  & $4.36\pm0.18$ \\
$w2$   & 1991 & 60533.0952 & $15.060\pm0.032$  & $5.06\pm0.15$ \\
&&&\\
$w2$   & 1991 & 60533.6903 & $15.519\pm0.048$  & $3.32\pm0.15$ \\
$w2$   & 1991 & 60533.7526 & $15.437\pm0.044$  & $3.58\pm0.14$ \\
$w2$   & 1991 & 60533.9462 & $15.050\pm0.037$  & $5.11\pm0.18$ \\

&&&\\
$u$    & 3442 & 60546.4252 & $15.922\pm0.054$  & $1.51\pm0.08$ \\
$u$    & 3442 & 60546.4860 & $15.984\pm0.045$  & $1.42\pm0.06$ \\
$w1$   & 2486 & 60546.4239 & $15.468\pm0.048$  & $2.59\pm0.11$ \\
$w1$   & 2486 & 60546.4837 & $15.474\pm0.039$  & $2.57\pm0.09$ \\
$m2$   & 2221 & 60546.4217 & $15.229\pm0.047$  & $3.76\pm0.16$  \\
$m2$   & 2221 & 60546.4800 & $15.317\pm0.039$  & $3.46\pm0.13$ \\
$w2$   & 1991 & 60546.4273 & $15.183\pm0.037$  & $4.52\pm0.15$ \\
$w2$   & 1991 & 60546.4897 & $15.362\pm0.032$  & $3.83\pm0.11$ \\\hline
\multicolumn{4}{l}{$^*$Data not dereddened.}
 \end{tabular}
\end{table*}

\section{Description of spectra}

\subsection{Day 8.58}

The day~8.58 spectrum shows the coronal line [\ion{Si}{X}]
1.4305\mic\ to be extraordinarily strong
(see Fig.~\ref{data}). 
The dereddened flux is
$f=[1.22\pm.01]\times10^{-15}$~W~m$^{-2}$, which
translates to $\sim95$\Lsun. This is significantly greater than 
the [\ion{Si}{X}] 1.43\mic\ flux in U~Sco at the same time 
post eruption \citep[$\sim1.5$\Lsun;]
[see Fig.~\ref{lmc-usco} below]{evans23},
and much larger than the highest measured [\ion{Si}{X}] flux
in the nova V1674~Her \citep[$\sim23$\Lsun;][]{woodward21}.
The line is clearly
a very powerful coolant of the coronal gas. 
Following \cite{woodward21}, $kT_*$ defines the temperature
$T_*$ at which half the photons emitted by the corresponding
blackbody are capable of ionising the lower ion
($kT_*\simeq0.426~\times$~ionisation potential).
For [\ion{Si}{X}], $kT_*=146$~eV, and as the temperature
of the X-ray-emitting gas was $kT_{\rm Swift}\simeq49$~eV, it
seems unlikely that the Si was photo-ionised.

[\ion{Si}{IX}] 1.5560\mic\ ($kT_*=127$~eV) 
is often seen in the wavelength range covered by the 
Magellan data, but it is not present, to a
$3\sigma$ limit of $\sim1.9\times10^{-16}$~W~m$^{-2}$. 
The presence of [\ion{Si}{X}] 1.4305\mic, coupled
with the absence of [\ion{Si}{IX}] 1.5600\mic,
may be used to estimate the temperature of
the coronal gas, using \citep[e.g.,][]{greenhouse90}
\[ \frac{f([\ion{Si}{IX}])}{f([\ion{Si}{X}])} =
\frac{n(\ion{Si}{IX})}{n(\ion{Si}{X})} 
\frac{\lambda([\ion{Si}{X}])}{\lambda([\ion{Si}{IX}])}
\frac{\Upsilon(\ion{Si}{IX})}{\Upsilon(\ion{Si}{X})}
\frac{g_l(\ion{Si}{X})}{g_l(\ion{Si}{IX})} \:\:,\]
where $n$ is the number density of each ion, $\Upsilon$
is the effective collision strength (i.e. the collisional
strength averaged over a thermal electron distribution),
and $g_l$ is the statistical weight of the lower level.
The apparent absence of [\ion{Si}{IX}] 1.5600\mic\
suggests that the temperature of the coronal gas was  
$T_{\rm cor}\gtsimeq10^{6.45}$~K ($\sim3\times10^6$~K). We have
used the values of $\Upsilon$ from the IRON Project online 
database\footnote{http://cdsweb.u-strasbg.fr/tipbase/home.html}
\citep{hummer93,badnell06}; we have taken the 
temperature-dependence of $\Upsilon$ for 
[\ion{Si}{IX}], and the value at $10^5$~K 
(the highest temperature available) for [\ion{Si}{IX}].
We have also used ionisation data from \cite{arnaud85}.

The time by which the NIR coronal lines appeared in LMC68
following its 2024 eruption 
(i.e., $t < 8.58$~d) is comparable to that seen
in other ``fast'' novae\footnote{Novae whose visual light curves
decline by 2~magnitudes on a time-scale $\ltsimeq20$~days
\citep{warner12}.}
\citep[9.41~d and 11.51~d in U~Sco and 
V1674~Her respectively;][]{woodward21,evans23}.
This suggests that the early ($\ltsimeq10$~days) appearance
of NIR coronal lines during nova eruptions 
(both classical and recurrent) is not unusual in fast novae.
However, the coronal phase in slower novae may occur much
later \citep[see the compilation in Table~VIII of][] 
{benjamin90}. 

\begin{table*}
\caption{Dereddened line fluxes. Upper limits are $3\sigma$.
\label{fire}}
 \begin{tabular}{ccccc} \hline
$\lambda_{\rm c}$ (observed)  &  ID & $\lambda$ ID & \multicolumn{2}{c}{Dereddened Flux} \\\cline{4-5}
\rule{0pt}{3ex}  (\mic)     &     & (\mic)  & Day 8.58 ($10^{-17}$~W~m$^{-2}$)  & 
Day 22.49 ($10^{-18}$~W~m$^{-2}$)  \\ \hline
1.010 &  \ion{He}{II} 5-4 & 1.01264  & $10.39\pm0.06$ & $1.97\pm0.12$\\
1.083   &  \ion{He}{I}      & 1.08332  & $24.66\pm0.02$& $<0.02$\\
1.163   &  \ion{He}{II} 7-5 & 1.16296  & $2.09\pm0.32$ & $1.50\pm0.07$\\
1.285  &  Pa$\,\beta$       & 1.28216  & $3.02\pm0.14$ & $<0.02$\\
1.431  &  [\ion{Si}{X}]      & 1.43049  & $122\pm10$ &
$13.4\pm0.2$\\\hline
 \end{tabular}
\end{table*}

Also present, but far weaker, are Pa$\,\beta$ 
1.2818\mic, \ion{He}{I} 1.0833\mic, \ion{He}{II} 5-4 1.0126\mic,
and \ion{He}{II} 7-5 1.163\mic. The dereddened flux ratio 
$f(\mbox{\ion{He}{II} 7-5})/f(\mbox{\ion{He}{II} 5-4})$ is 
approximately 0.2.
Assuming that Case~B \citep{storey95} applies to the
\ion{He}{II}-bearing gas, this flux ratio is consistent
with a range of electron densities 
($10^3<n_e{\mbox{~(cm$^{-3}$)}}<10^{10}$) and 
electron temperatures
($5000<T_e\mbox{~(K)}<10^5$).

\begin{figure}
 \includegraphics[width=8cm,keepaspectratio]{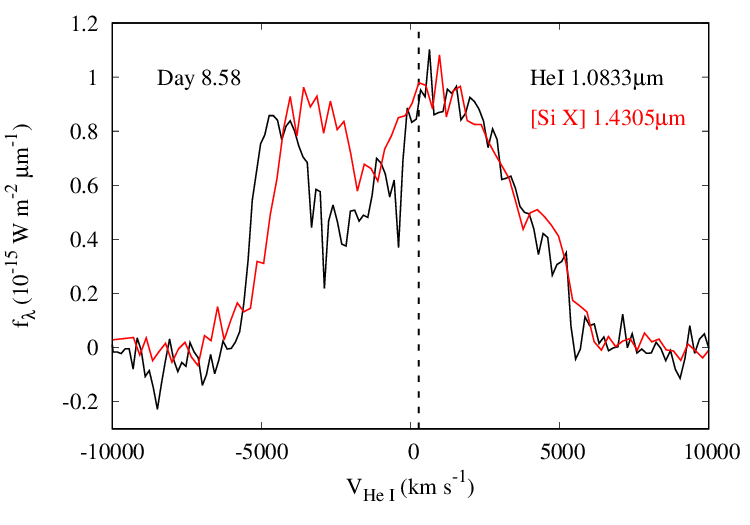}
 \caption{Line profile of the \ion{He}{I} 1.083\mic\ 
 and [\ion{Si}{X}] 1.4305\mic\ features in velocity space
 on day 8.58 from the Magellan spectrum. 
 The vertical dotted line is the systemic velocity
 of the LMC.\label{HeI}}
\end{figure}

Both [\ion{Si}{X}] and \ion{He}{I} are double-peaked
on day 8.58, with half width at zero intensity (HWZI)
$\sim5000$\vunit\ and separation between peaks of
$\sim5000$\vunit\ (see Fig.~\ref{HeI});
this is broadly similar to the H$\,\alpha$ profile 
on day~4.7 \citep{shore24}, although the latter
had a narrow central portion in the velocity range 
--900\vunit\ to 1780\vunit. While the profiles
of [\ion{Si}{X}] and \ion{He}{I} are similar, 
it is unlikely that they arise in the same region of
the ejecta, for reasons discussed below.
The high velocity and the profile suggest that, as in
\us, they arise in 
collimated outflows ejected orthogonal to the 
binary plane \citep{evans23}.
The high inclination of the binary \citep{kuin20}
suggests that the ejected velocity is much higher than this.
The critical electron density for the [\ion{Si}{X}] 1.4305\mic\
transition (above which the upper level is
collisionally rather than radiatively de-excited) 
is $n_{\rm crit} \simeq 2.76\times10^5\:T_e^{1/2}$~cm$^{-3}$,
where $T_e$ is the electron temperature. From the silicon
coronal lines ([\ion{Si}{X}] 1.4305\mic, [\ion{Si}{IX}]
1.5600\mic), we determined 
that $T_{\rm cor}\gtsimeq3\times10^6$~K, so that 
$n_{\rm crit} \gtsimeq 4.9\times10^8$~cm$^{-3}$
if $T_{\rm cor}=T_e$.
The [\ion{Si}{X}] 1.4305\mic\ and \ion{He}{II} lines 
can be co-spatial if the 
electron density in this region is $\ltsimeq10^8$~cm$^{-3}$.

\begin{figure}
 \includegraphics[width=8cm,keepaspectratio]{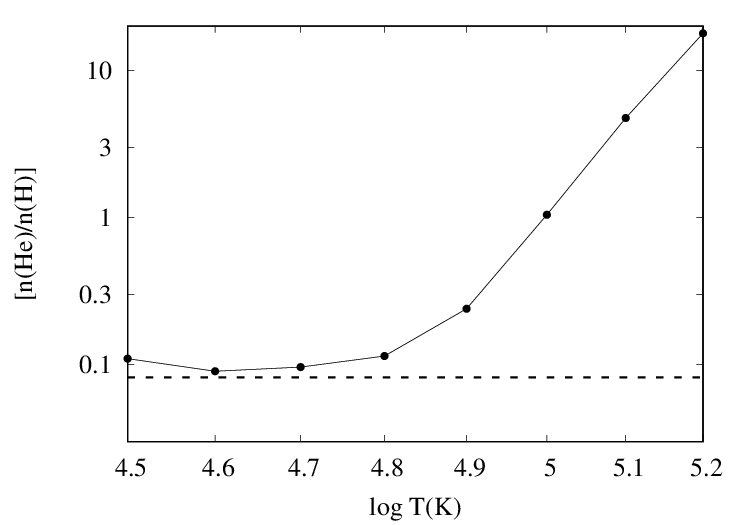}
 \caption{The He/H abundance ratio as function of temperature
 from, Equation~(\ref{HeH}).
 Horizontal dotted line is solar He/H ratio \citep*{asplund21}.
 \label{He-abund}}
\end{figure}

Both Pa$\,\beta$ and \ion{He}{II} 5--4 are present in the 
day~8.58 data. For both lines the upper level is $n=5$, 
so the expected flux ratio has the simple form (assuming that both
lines are optically thin):
\[ \frac{f(\mbox{Pa}\beta)}{f(\mbox{\ion{He}{II}})} = 
 \frac{n(\mbox{H})}{n(\mbox{\ion{He}{II}})} \:\: 
 \exp \left ( -\frac{3hRc}{25kT}  \right ) \:\:  
 \frac{A(\mbox{Pa}\beta)}{A(\mbox{\ion{He}{II}})} \:\:
 \frac{\lambda(\mbox{\ion{He}{II}})}{\lambda(\mbox{Pa}\beta)}  \] 
so that 
\begin{equation}
 \frac{n(\mbox{He})}{n(\mbox{H})}  =  0.1384 \:\:\:
\exp \left ( -\frac{1.894\times10^4}{T\mbox{~(K)}}  \right )
 \frac{n(\mbox{He})}{n(\mbox{\ion{He}{II}})} \:\:. \label{HeH}
 \end{equation}
Here $R$ is the Rydberg constant, and $A$ is the Einstein
spontaneous emission coefficient
($A=2.201\times10^6$~s$^{-1}$ for Pa$\beta$,
$4.321\times10^{7}$~s$^{-1}$ for \ion{He}{ii} 5--4). The He/H
ratio by number is therefore a simple function of the 
temperature $T$ (see Fig.~\ref{He-abund}), the 
temperature-dependence of the ratio 
$n(\mbox{He})/n(\mbox{\ion{He}{II})}$ being 
given in \cite{arnaud85}. 
If $n$(He)/$n$(H) $\simeq1$, as found by \cite{shore91}
for the 1990 eruption, and this ratio also applies to the 2024
eruption, the implication is that \ion{He}{II} originates
in a region at $\sim10^5$~K  (see Fig.~\ref{He-abund}), 
considerably cooler than, and distinct from, the hotter 
($T\simeq10^{6.45}$~K) [\ion{Si}{x}] 1.4305\mic\ emitting region.
Indeed nearly all He is in the form  
\ion{He}{III} above $10^5$~K \citep{arnaud85}.

The dereddended Swift UVOT data for this epoch are consistent
with a black body with a temperature of $19700\pm2030$~K and
luminosity $1.8[\pm1.0]\times10^3$\Lsun.

\subsection{Day 22.49}
On day~22.49, the [\ion{Si}{X}] 1.4305\,\mic\ line 
was detected, 
but its flux had decreased by a factor $\sim100$, to 
$[1.34\pm0.02]\times10^{-17}$~W~m$^{-2}$.
Also greatly diminished, but still present, were \ion{He}{II} 
1.0126\mic\ and \ion{He}{II} 1.163\mic.
However, \ion{He}{I} 1.0833\mic\ and Pa$\,\beta$
were no longer detected.

\begin{figure}
 \includegraphics[width=8cm,keepaspectratio]{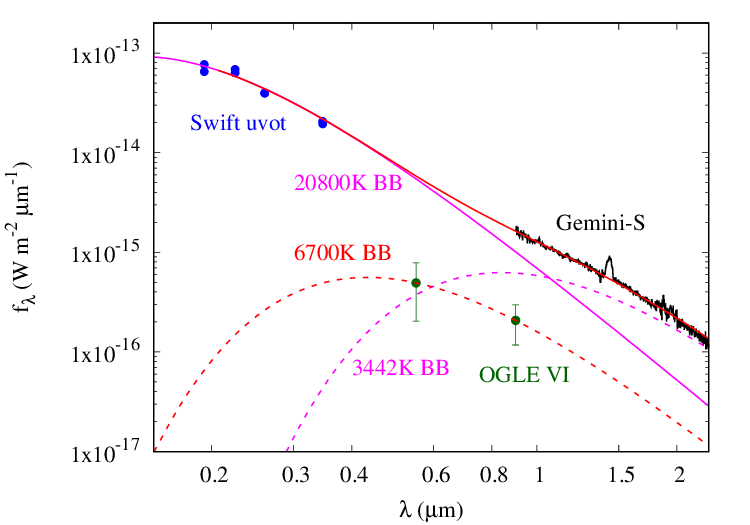}
 \caption{The UV-IR flux distribution of LMC68 on day 22.49.
 All data dereddened. Blue circles, Swift UVOT data; black line:
 Gemini-S NIR data. 
 Full magenta curve, 20800~K black body fit to Swift data.
 Dotted magenta curve, 3442~K  black body fit to
 excess as described in text. Full red curve is sum of
 20800~K and 3442~K black bodies.
 Dotted red curve, 6700~K black body fit to OGLE data (green);
 see discussion in Section~\ref{system}.
 \label{cont23}}
\end{figure}

The dereddened Swift UVOT data for day~22 are
consistent with a black body with temperature 
$20800\pm2400$~K, and 
luminosity $\simeq[1.5\pm0.9]\times10^3$\Lsun\
(see Fig.~\ref{cont23}) These values are not significantly
different from those apparent on day~8.58. 
The UVOT-derived temperatures on days~8.58
and 22.49 ($\simeq2\times10^4$~K) and
bolometric luminosity ($\simeq1.5\times10^3$\Lsun)
are remarkably similar to the same parameters obtained
for \us\ during its 2022 eruption \citep{evans23}.
The corresponding ``black body angular diameter''
is $\sim0.05$\Rsun. As in the case of \us, we explore the 
possibility that this emission is due to irradiation
of the secondary by the hot WD. For a secondary star of mass
$\sim1$\Msun, the Roche lobe radius is
$\sim0.2$\Rsun. The formalism of \cite*{kovetz88} gives the
irradiated temperature as $\sim10^5$~K and luminosity 
$\sim2\times10^4$\Lsun, if the 
WD was radiating at the Eddington luminosity.
It seems unlikely, therefore, that this emission is
the result of irradiation of the
secondary by the hot WD. 

As shown in Fig.~\ref{cont23}, the extrapolation
of this black body to longer wavelengths reveals an excess
in the NIR. A NIR wavelength excess was also evident in the
outburst NIR spectrum of \us\ \citep{evans23},
which was attributed to free-free and free-bound emission.
In the case of LMC68, however, the excess is not fitted by
free-free and free-bound emission, but is consistent
with black body emission at $\sim3400$~K and luminosity $\sim63$\Lsun\
(see Fig.~\ref{cont23}).

\subsection{Comparison with U Sco}

\begin{figure*}
 \includegraphics[width=8cm]{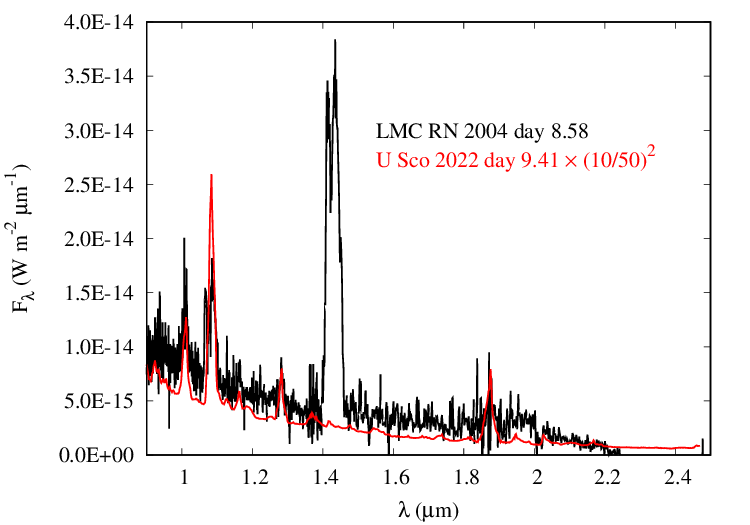}
 \includegraphics[width=8cm]{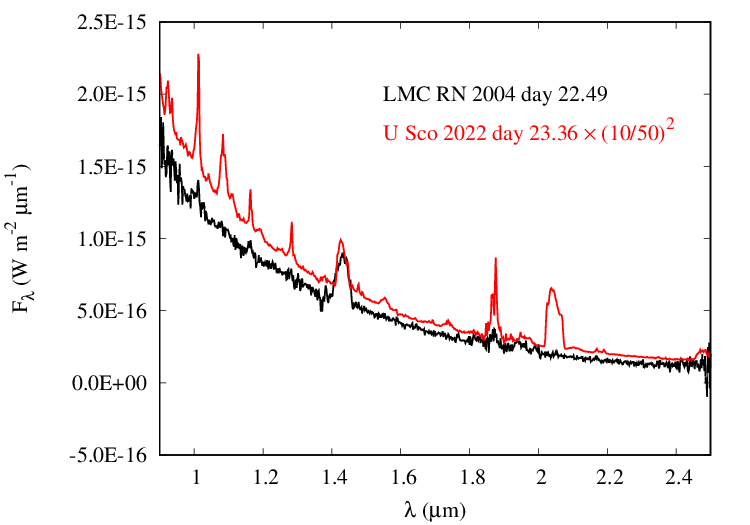}
 \caption{Comparison of the NIR spectra of LMC68
 and \us\ \citep{evans23}, obtained at similar 
 intervals after eruption.\label{lmc-usco}}
\end{figure*}

With the limited amount of LMC68 data at our disposal, 
we compare LMC68 with the Galactic RN \us\ at equivalent 
stages of their outbursts.
The latter has had RN eruptions in 1863, (1873, 1884, 1894),
1906, 1917, (1927), 1936, 1945, (1955), 1969, 1979, 1987, 1999,
2010, (2016), and 2022, dates in brackets being probable eruptions
which were missed \citep[see summary in][]{evans23}. The mean time
between eruptions for \us\ is $\sim10$~yr, of the same order
as the 4-yr recurrence time for LMC68.

Fortuitously, the NIR data reported here were obtained at 
similar times after eruption as some of the NIR 
spectroscopy of the 2022 eruption of \us\ \citep{evans23}. 
To compare the data we have scaled the \us\ spectra to 
the LMC by applying a factor $[10/50]^2$, where the 
distance of \us\ is taken to be 10~kpc. The comparison
is shown in Fig.~\ref{lmc-usco}. Given the crude nature of
the scaling, the continuum levels
are strikingly similar, but there are clear differences
between the two objects (the day numbers below are for LMC68):
\begin{enumerate}
 \item Day 8.58: the earlier spectra are broadly similar,
 in terms of the shape of the continuum and the emission lines
 present. The glaring exception is the extraordinary strength
 of [\ion{Si}{X}] 1.4305\mic\ in LMC68. Other
 than this line there is no clear evidence for any metal
 lines in the NIR spectrum.
 \item Day 22.49: the later spectra are very different.
 The NIR continuum of \us\ is much
 bluer than that of LMC68, essentially because there 
 is a greater contribution from the hot black body in the 
 former. The emission lines in LMC68 are 
 (with the exception of [\ion{Si}{X}] 1.4305\mic)
 weak or non-existent, in contrast to \us. 
 The absence of \ion{He}{I} 1.0833\mic\
 in LMC68 is intriguing, as the strength of this line
 in \us\ went 
 through a flux minimum on day 23.36 \citep{evans23}.
\end{enumerate}

\section{Where are the metal lines?}

A striking feature of the LMC68 NIR spectra reported 
here, compared to the apparently similar system \us,
is the absence of any metal lines, with the 
exception of the [\ion{Si}{X}] coronal line at
$\lambda=1.4305$\mic. In addition to H and He lines, 
lines of aluminium, sulphur and calcium are generally present
(e.g [\ion{Al}{IX}] 2.04\mic, [\ion{S}{IX}] 1.25\mic, 
[\ion{Ca}{VIII}] 2.302\mic). 
In particular, the Al and the S lines are usually present
along with the [\ion{Si}{X}] 1.43\mic\
line, all the ions involved  having similar ionisation
potentials, in the range 300--400~eV
\citep[e.g., Table~2 in][and \citeauthor{evans23} 
\citeyear{evans23}]{woodward21}.
In general, the [\ion{Al}{IX}] 2.04\mic\ line is very strong, 
often as strong as [\ion{Si}{X}] 1.43\mic\
(see, e.g., the NIR spectrum of \us\ in
the right panel of Fig.~\ref{lmc-usco}).
As noted above,
an optical spectrum on day~4.7 of the 2024 eruption
\citep{shore24} was similarly devoid of metal lines.

In addition to the above NIR coronal lines, a number of 
other, weaker, coronal lines are often seen during nova
eruptions, for example [\ion{P}{VIII}] 1.7361\mic,
[\ion{Cr}{XI}] 1.5503\mic, [\ion{Ti}{VI}] 1.715\mic, and
[\ion{Mn}{XIV}] 2.092\mic\
 \citep{wagner96,raj15,gehrz18,kumar22,rudy24}. 

The issue is not solely an absence of coronal lines.
NIR spectra of the 2006 eruption of the RN RS~Oph
have (in addition to coronal lines) \ion{O}{I}
1.1287, 1.3164\mic, \ion{C}{I} 1.1748\mic, and \ion{Fe}{ii}
1.6872\mic\ \citep*{banerjee09}; note that it is possible 
that these lines may have contributions from the RG wind.

Metal lines {\em have} been seen in outburst spectra
of LMC68. Spectra of the 1990 eruption obtained with the 
International Ultraviolet Explorer (IUE) satellite 
\citep{shore91} show \ion{N}{V} 1240\AA, \ion{Si}{IV} 1400\AA,
and \ion{C}{IV} 1500\AA, all of which weakened
considerably over the first $\sim10$~days of the eruption.
Swift UVOT grism spectra obtained during the first 10~days
of the 2016 eruption \citep{kuin20} showed \ion{C}{iii}]
1909\AA\ and \ion{O}{iii} 3133\AA; these emission lines 
were rather weak.

While the NIR properties of LMC68 may merely 
be a consequence of its extreme nature, we consider
below other possible reasons for the metal-deficient
nature of the 2024 LMC68 ejecta.

\subsection{Electron density}

Might the electron density in the coronal region
have been so high as to permit the collisional de-excitation
of the transition upper levels? This is of course possible
but can not explain the absence of other lines 
(e.g., \ion{O}{I} 1.1287, 1.3164\mic) commonly
seen in the evolution of the NIR spectra of RNe.

\subsection{TNR when the secondary is metal poor}

LMC68 differs from Galactic RNe in that
the metallicity of the secondary is likely to be typical
of the LMC. The metallicity of LMC RGs is 
$\mbox{[Fe/H]}\simeq-0.6$ \citep*{cole00}. Nova explosions
occurring in binaries in which the secondary has 
less-than-Solar metallicity have been discussed by 
\cite{chen19}, \cite{starrfield00}, and \cite{jose07}.
As discussed by \cite{starrfield12}, the accretion of 
low-metallicity material on to the WD results in a more violent
explosion. This is because the key driver of the TNR is
the \nucl{12}{C}$(p,\gamma)$\nucl{13}{N} reaction, so the
lower the metallicity, the more accreted material is required to
power the outburst, the greater the pressure at the base of the 
accreted envelope and the greater the strength of the eruption.

Even though the secondary is likely metal-deficient, the 
material accreted on the surface of the WD is still processed
through the TNR, along with material dredged up from the WD
(although the extent to which the latter would
occur in an ultra-short recurrence period RN is not clear).
The composition of the ejecta will therefore reflect this.
Indeed, in view of the likely more energetic eruption,
metals beyond Ca (the usual end-point of the TNR in nova
explosions), are expected in a nova arising in a binary
with a metal-deficient secondary \citep{jose07}. 
Although those authors considered only cases of 
{\em extreme} metal deficiency (e.g., solar/$2\times10^5$),
their models suggest that C, N, O, Si, Al, S, P are all 
expected to be enhanced relative to solar, whereas Mg 
and Ca are expected to
be {\em deficient} relative to solar.

Thus, the metal-deficiency of the secondary can not by itself
account for the lack of metal lines in the NIR.

\subsection{Collisional ionisation}

The X-ray source in LMC68 was never hot enough
to ionise the gas to the extent that coronal lines would
be present ($kT_{\rm Swift}$ was less than
the ionisation potential for the lower ionisation states). 
The alternative is that the \ion{Si}{X} is
produced by shocks. The fact that the secondary is
not a RG means that there is little or no 
wind with which the nova ejecta can interact, 
but there might be interaction between ejected material
moving at different velocities.
It is known that velocities up to 5000\vunit\ are
present \citep[see above and][]{shore24}.
\citeauthor{shore24} also note the presence of 
narrower central features in the interval 
$-900$\vunit\ to 1780\vunit, suggesting that 
there are regions of the ejecta with relative velocities
$V_{\rm rel}$ in excess of 1000\vunit. The RMS velocity 
$V_{\rm rms}$ of a gas, even at $10^6$~K, is at most 170\vunit,
irrespective of the composition of the gas. Thus 
$V_{\rm rel}/V_{\rm rms}\gg1$, so the gas is strongly shocked,
and its temperature is
\[ T_{\rm shock} \simeq \frac{3\mu{m_{\rm H}}}{16k} \:\: V_{\rm rel}^2 \:\:,\]
where $\mu$ is the mean atomic weight of the gas,
$m_{\rm H}$ is the mass of a H atom, and
$V_{\rm rel}$ is the relative velocity 
(likely $\gtsimeq1000$\vunit). Thus 
$kT_{\rm shock}\sim2$~keV, more than adequate to
account for both the presence of high-ionisation 
species and the high temperature implied by the Si lines
(at least $10^{6.45}$~K, $\sim3\times10^6$~K, $kT\gtsimeq270$~eV).
Interestingly, this is significantly higher than 
the coronal gas temperature inferred in other novae, e.g.,
$10^{5.5}$~K \citep[several novae;][]{greenhouse90}, 
$10^{6.2}$~K \citep[RS~Oph;][]{banerjee09},
$10^{5.6}$~K \citep[V1674~Her;][]{woodward21},
$10^{6.0}$~K \citep[V3890~Sgr;][]{evans22},
$10^{5.8}$~K \citep[\us;][]{evans23},
and is possibly the highest coronal temperature 
recorded in a nova eruption.

\begin{table}
\caption{Properties of the highest stages of ionisation
discussed in this paper. See text for definitions of
$T_*$ and $T_{1/2}$.\label{tplus}}
 \begin{tabular}{ccccc} \hline
 Ion          & $\lambda$ & IP$^\dag$  & $kT_*$ & $kT_{1/2}$ \\
              &  (\mic)   & (eV)& (eV)  & (eV) \\ \hline
 \fion{S}{ix} & 1.2523 & 329 & 137 & 278 \\
 \fion{P}{viii}& 1.375,1.7361 & 264 & 110 & 223 \\
  \fion{Si}{x} & 1.4305 & 351 & 146 & 297 \\ 
 \fion{Si}{xi} & 1.9320 & 401 & 167 & 339 \\
\fion{Al}{ix} & 2.0400 & 285 & 119 & 241\\
\fion{Ca}{viii} & 2.3211 & 127 & 53 & 107\\ \hline
\multicolumn{4}{l}{$^\dag$Ionisation potential of lower stage.}
 \end{tabular}
\end{table}

The higher coronal temperature in LMC68 might offer a clue
to the absence of metal lines in the NIR. 
The high temperature of the coronal gas 
leads to {\em collisional ionisation}
of ions, so that the abundances of neutral species 
(such as \ion{C}{I}, \ion{O}{I}) are negligible, while 
species normally appearing in IR coronal spectra
(such as [\ion{Ca}{VIII}]) are ionised to even 
higher stages. Thus, the fractional abundances of
the ions listed in Table~2 of \cite{woodward21}
(and itemised throughout the present paper)
all peak at temperatures below $10^{6.3}$~K,
their abundances diminishing at higher temperatures
\citep[see the tabulations in][]{arnaud85}.

For the species discussed in this work, the
ionisation potentials of the ions that
have higher stages of ionisation 
are listed in Table~\ref{tplus}.
These are generally less than that of \ion{Si}{IX}
(351.1~eV), and so lower ionisation stages of these
ions are unlikely to exist in a gas having temperature 
$\gtsimeq10^{6.45}$~K; they have been ionised to higher
stages by collisional ionisation. By analogy with the 
definition of $kT_*$, we define $T_{1/2}$ as the 
temperature of the Maxwellian gas in which at least 
half of the electrons are capable of ionising the lower
ionisation stage ($kT_{1/2}\simeq0.845~\times$ ionisation
potential). Values of both $kT_*$ and $kT_{1/2}$
are listed in Table~\ref{tplus}. While the X-ray
source is incapable of the necessary ionisation,
a gas at $kT\gtsimeq270$~eV is certainly capable; 
note that this value is generally comparable with, or
exceeds, the $kT_{1/2}$ values in Table~\ref{tplus}.

The absence of almost all of the commonly observed
metal lines in the NIR spectrum
of nova LMC68 may therefore be a consequence of the 
unusually high temperature of the coronal gas,
compounded by the expected underabundance of
Mg and Ca in a TNR from a metal-deficient fuel.
It may also account for the change in the NIR spectrum
we observed between days~8.58 and 22.49. Furthermore,
this change may also be consistent with the decline in
the ultraviolet line flux reported by \cite{shore91}
during the first week of the 1990 eruption.

\section{Conclusion}

We have presented the first NIR spectra of an extragalactic
recurrent nova (OGLE 2016-NOVA-1, LMC V1341, LMCN 1968-12a)
in eruption. The spectra, obtained 8.58 and 22.49~days
after the eruption, are remarkable in that they are
almost devoid of metal lines, with the sole exception 
of \fion{Si}{x} 1.4305\mic\ line, which was
extraordinarily strong on day~8.58.

We suggest that the absence of other metal lines in the
NIR is a consequence of (a)~the exceptionally high
temperature of the coronal gas, resulting in the 
collisional ionisation of ions normally seen 
during the coronal phase of novae to even higher states
of ionisation, so that the abundances of
the states normally observed in the coronal
gas are almost negligible, and (b)~the underabundance
of Mg and Ca in a metal-deficient gas that has been 
processed through a TNR.
However, to better quantify this conclusion, it 
will be necessary to model
the products of a RN TNR for systems having secondary
stars with LMC abundances.

\section*{Data availability}

The raw infrared data in this paper are available 
from the Gemini Observatory
Archive, https://archive.gemini.edu/.

The Magellan data will be made available on reasonable
request to the first author.

The Swift data are available from
https://www.swift.ac.uk/swift\_live/ and 
https://heasarc.gsfc.nasa.gov/cgi-bin/W3Browse/ \\ w3browse.pl~.

VISTA data are available at the European 
Southern Observatory Science Archive 
http://archive.eso.org/cms.html~.

\section*{Acknowledgements}
We thank the referee for their helpful
comments on an earlier version of this paper.

The Gemini observations described in this paper were made 
possible through the award of Director's Discretionary Time
GS-2024B-DD-104.
The international Gemini Observatory is a program of 
NSF's NOIRLab, which is managed by the Association of Universities 
for Research in Astronomy (AURA) under a cooperative agreement 
with the National Science Foundation, on behalf of the Gemini 
Observatory partnership: the National Science Foundation 
(United States), National Research Council (Canada), 
Agencia Nacional de Investigaci\'{o}n y Desarrollo (Chile), 
Ministerio de Ciencia, Tecnolog\'{i}a e Innovaci\'{o}n (Argentina), 
Minist\'{e}rio da Ci\^{e}ncia, Tecnologia, 
Inova\c{c}\~{o}es e Comunica\c{c}\~{o}es (Brazil), 
and Korea Astronomy and Space Science Institute (Republic of Korea).

AP acknowledges prior support by a Carnegie Fellowship 
through which the Magellan time observing time was awarded.
KLP acknowldges funding from the UK Space Agency.
SS acknowledges partial support from a NASA Emerging Worlds
grant to ASU (80NSSC22K0361) as well as support from his
ASU Regents' Professorship.



%


\bsp	
\label{lastpage}
\end{document}